  \documentclass[conference]{IEEEtran}
  \usepackage{graphicx}
  \usepackage[T1]{fontenc}
  \usepackage{cite}

\ifCLASSINFOpdf
\else
\fi
\usepackage{amsmath, amsthm, amssymb}
\usepackage[cmintegrals]{newtxmath}

\makeatletter
\def\abstract{\normalfont
    \if@twocolumn
      \noindent\@IEEEabskeysecsize\bfseries\textit{\abstractname}---\relax
    \else
      \bgroup\par\addvspace{0.5\baselineskip}\centering\vspace{-1.78ex}\@IEEEabskeysecsize\textbf{\abstractname}\par\addvspace{0.5\baselineskip}\egroup\quotation\@IEEEabskeysecsize
    \fi\@IEEEgobbleleadPARNLSP}
\def\endabstract{\relax\ifCLASSOPTIONconference\vspace{0ex}\else\vspace{1.34ex}\fi\par\if@twocolumn\else\endquotation\fi
    \normalfont\normalsize}

\def\IEEEkeywordsname{Keywords}
\def\IEEEkeywords{\normalfont
    \if@twocolumn
      \@IEEEabskeysecsize\bfseries\itshape\IEEEkeywordsname---\relax
    \else
      \bgroup\par\addvspace{0.5\baselineskip}\centering\@IEEEabskeysecsize\textbf{\IEEEkeywordsname}\par\addvspace{0.5\baselineskip}\egroup\quotation\@IEEEabskeysecsize
    \fi\@IEEEgobbleleadPARNLSP}
\def\endIEEEkeywords{\relax\ifCLASSOPTIONtechnote\vspace{1.34ex}\else\vspace{0.67ex}\fi
    \par\if@twocolumn\else\endquotation\fi%
    \normalfont\normalsize}

\makeatother

\usepackage{url,multirow,morefloats,floatflt,cancel,tfrupee}
\makeatletter

\AtBeginDocument{\@ifpackageloaded{textcomp}{}{\usepackage{textcomp}}}
\makeatother
\usepackage{colortbl}
\usepackage{xcolor}
\usepackage{pifont}
\usepackage[nointegrals]{wasysym}
\makeatletter

\def\mcWidth#1{\csname TY@F#1\endcsname+\tabcolsep}

\def\cAlignHack{\rightskip\@flushglue\leftskip\@flushglue\parindent\z@\parfillskip\z@skip}
\def\rAlignHack{\rightskip\z@skip\leftskip\@flushglue \parindent\z@\parfillskip\z@skip}

\usepackage{ifxetex}
\ifxetex\else\if@twocolumn\usepackage{dblfloatfix}\fi\fi

\AtBeginDocument{
\expandafter\ifx\csname eqalign\endcsname\relax
\def\eqalign#1{\null\vcenter{\def\\{\cr}\openup\jot\m@th
  \ialign{\strut$\displaystyle{##}$\hfil&$\displaystyle{{}##}$\hfil
      \crcr#1\crcr}}\,}
\fi
}

\AtBeginDocument{%
  \@ifpackageloaded{endfloat}%
   {\renewcommand\efloat@iwrite[1]{\immediate\expandafter\protected@write\csname efloat@post#1\endcsname{}}}{}%
}%

\let\lt=<
\let\gt=>
\def\processVert{\ifmmode|\else\textbar\fi}

\@ifundefined{subparagraph}{
\def\subparagraph{\@startsection{paragraph}{5}{2\parindent}{0ex plus 0.1ex minus 0.1ex}%
{0ex}{\normalfont\small\itshape}}%
}{}

\newcommand\role[1]{\unskip}
\newcommand\aucollab[1]{\unskip}
  
\@ifundefined{tsGraphicsScaleX}{\gdef\tsGraphicsScaleX{1}}{}
\@ifundefined{tsGraphicsScaleY}{\gdef\tsGraphicsScaleY{.9}}{}
\def\checkGraphicsWidth{\ifdim\Gin@nat@width>\linewidth
	\tsGraphicsScaleX\linewidth\else\Gin@nat@width\fi}

\def\checkGraphicsHeight{\ifdim\Gin@nat@height>.9\textheight
	\tsGraphicsScaleY\textheight\else\Gin@nat@height\fi}

\def\fixFloatSize#1{}%\@ifundefined{processdelayedfloats}{\setbox0=\hbox{\includegraphics{#1}}\ifnum\wd0<\columnwidth\relax\renewenvironment{figure*}{\begin{figure}}{\end{figure}}\fi}{}}
\let\ts@includegraphics\includegraphics

\def\inlinegraphic[#1]#2{{\edef\@tempa{#1}\edef\baseline@shift{\ifx\@tempa\@empty0\else#1\fi}\edef\tempZ{\the\numexpr(\numexpr(\baseline@shift*\f@size/100))}\protect\raisebox{\tempZ pt}{\ts@includegraphics{#2}}}}

\AtBeginDocument{\def\includegraphics{\@ifnextchar[{\ts@includegraphics}{\ts@includegraphics[width=\checkGraphicsWidth,height=\checkGraphicsHeight,keepaspectratio]}}}

\def\URL#1#2{\@ifundefined{href}{#2}{\href{#1}{#2}}}

\def\UrlOrds{\do\*\do\-\do\~\do\'\do\"\do\-}%
\g@addto@macro{\UrlBreaks}{\UrlOrds}

\@ifundefined{quoteAttrib}
	{}
	{}

\@ifundefined{titlequoteAttrib}
	{}{}

\newenvironment{title-quote}
	{\list{}{\fontsize{10pt}{12pt}\selectfont\leftmargin.5in\itshape\rightmargin\leftmargin}%
  \item\relax}
  {\endlist}

\makeatother

\makeatletter
\AtBeginDocument{\@ifpackageloaded{longtable}{%
\def\LT@makecaption#1#2#3{%
  \LT@mcol\LT@cols c{\hbox to\z@{\hss\parbox[t]\LTcapwidth{%
    \sbox\@tempboxa{#1{#2: } #3}%
    \ifdim\wd\@tempboxa>\hsize
      #1{#2: }\textsc{#3}%
    \else
      \hbox to\hsize{\hfil\box\@tempboxa\hfil}%
    \fi
    \endgraf\vskip\baselineskip}%
  \hss}}}
}{}}
\makeatother

   \makeatletter
  \def\fig@textbf{\textbf}
   \AtBeginDocument{\renewcommand\floatc@plain[2]{\setbox\@tempboxa\hbox{{\footnotesize#1.}\footnotesize\hskip.5em#2}%
    \ifdim\wd\@tempboxa>\hsize {\fig@textbf{\footnotesize#1.}}\footnotesize\hskip.5em#2\par
        \else\hbox to\hsize{\hfil\box\@tempboxa\hfil}\fi}}
    \makeatother
 \usepackage{float}
\newtheorem{theorem}{Theorem}
\newtheorem{lemma}{Lemma}
\newtheorem{Observation}{Observation}
\usepackage{enumitem}
\newlist{Properties}{enumerate}{2}
\setlist[Properties]{label=Property \arabic*.,itemindent=*}
\DeclareMathOperator*{\argmax}{arg\,max} 
\begin{document}

%
% paper title
% Titles are generally capitalized except for words such as a, an, and, as,
% at, but, by, for, in, nor, of, on, or, the, to and up, which are usually
% not capitalized unless they are the first or last word of the title.
% Linebreaks \\ can be used within to get better formatting as desired.
% Do not put math or special symbols in the title.

% conference papers do not typically use \thanks and this command
% is locked out in conference mode. If really needed, such as for
% the acknowledgment of grants, issue a \IEEEoverridecommandlockouts
% after \documentclass

        %\title{Edge Computing with UAV: Trajectory Optimization for Fairness in IoT Applications}
      %\title{Mid-air re-charging of UAVs using Wireless Power Transfer}
      %\title{UAV-enabled Wireless Power Transfer: the Case of Low-Power UAV Re-charging}
      %\title{In-air Charging of UAVs using Wireless Power Transfer}
       
    \title{Recharging of Flying Base Stations using Airborne RF Energy Sources}
     
      \author{
		\IEEEauthorblockN{Jahan~Hassan}
\IEEEauthorblockA{School of Engineering and Technology \\Central Queensland University\\ 
        Australia\\Email: j.hassan@cqu.edu.au}
        \vspace*{1pc}\and 
		\IEEEauthorblockN{Ayub~Bokani}
    \IEEEauthorblockA{School of Engineering and Technology \\Central Queensland University\\ 
        Australia\\Email: a.bokani@cqu.edu.au}
        \vspace*{1pc}\and 
		\IEEEauthorblockN{Salil~S. Kanhere}
\IEEEauthorblockA{School of Computer Science \\
and Engineering, University of \\
New South Wales, Australia\\
Email: salil.kanhere@unsw.edu.au} \vspace*{1pc}
}
  
% use for special paper notices
%\IEEEspecialpapernotice{(Invited Paper)}

% make the title area

\maketitle 
% As a general rule, do not put math, special symbols or citations
% in the abstract or keywords.

\begin{abstract}
This paper presents a new method for recharging flying base stations, carried by Unmanned Aerial Vehicles (UAVs), using wireless power transfer from dedicated, airborne, Radio Frequency (RF) energy sources. In particular, we study a system in which UAVs receive wireless power without being disrupted from their regular trajectory. The optimal placement of the energy sources are studied so as to maximize received power from the energy sources by the receiver UAVs flying with a linear trajectory over a square area. We find that for our studied scenario of two UAVs, if an even number of energy sources are used, placing them in the optimal locations maximizes the total received power, while achieving fairness among the UAVs. However, in the case of using an odd number of energy sources, we can either maximize the total received power, or achieve fairness, but not both at the same time. Numerical results show that placing the energy sources at the suggested optimal locations results in significant power gain compared to non-optimal placements.
\end{abstract}


% For peer review papers, you can put extra information on the cover
% page as needed:
% \ifCLASSOPTIONpeerreview
% \begin{center} \bfseries EDICS Category: 3-BBND \end{center}
% \fi
%
% For peerreview papers, this IEEEtran command inserts a page break and
% creates the second title. It will be ignored for other modes.
\IEEEpeerreviewmaketitle

\section{Introduction}

Recent advances in miniaturization, robotics, sensor technology and communications have revolutionized Unmanned Aerial Vehicles (UAVs) and  brought about their adoption in a wide range of applications. One such application is the use of UAVs carrying base station equipment acting as aerial base stations that can dynamically re-position themselves to improve coverage and capacity demands of existing networks \unskip~\cite{flyingdronebs,dronebsplacement,DBLP:journals/corr/abs-1809-01752,7974285,uav_wcomm, 7994915,dronebs2}. Such aerial base stations could supplement terrestrial infrastructure when it is overloaded or unavailable, as presented in the context of $5G$ networks in \unskip~\cite{5gtutorial, UAV5g}. While the majority of these proposals considered non-mobile UAVs hovering over a service area, some recent works \unskip~\cite{7974285, flyingdronebs, mabs} have argued for the use of flying (or cruising) aerial base stations wherein the UAVs continue to service ground nodes while in flight. The trajectory, i.e., the movement patterns of the aerial base stations is tailored so as to maximize network performance in the presence of geospatial variation in user demand, or to improve spectral efficiency. A prototype demonstrating the use of flying aerial base stations was developed recently by Eurecom \unskip~\cite{eurecom}.  

UAVs rely on an on-board battery for power, which limits their operational duration before recharging is required. Researchers have investigated power-efficient operations of UAVs to extend battery lifetime by reducing the energy consumed for communications (electronics) and mobility (mechanical), as summarized in \unskip~\cite{DBLP:journals/corr/abs-1809-01752}. Since extending the lifetime of the battery does not eliminate the need for recharging, a promising and parallel direction of research involves investigating ways for recharging UAVs to ensure service continuity. In particular, mechanisms for replenishing energy without disrupting the UAV's usual trajectory (in the case of flying UAVs) or deployed locations (in the case of non-mobile UAVs) where the UAV is not required to move to a different location to receive power, is essential for uninterrupted service provisioning. In this paper, we propose an architecture for recharging cruising UAVs using energy harvesting from received Radio Frequency (RF) transmitted by \textit{dedicated, non-mobile airborne} UAVs equipped with RF transmitters referred to as transmitter UAVs ($tUAV$s). In particular, we study the optimum placement of the $tUAV$s to maximize the received energy by the receiver UAVs ($rUAV$s). 

Researchers in \unskip~\cite{RF_UAV} have also explored RF energy harvesting for recharging UAVs. However, they rely on terrestrial energy sources for charging UAVs while we consider airborne chargers. As such, our approach can be used in a wide range of scenarios where deployment of terrestrial chargers may not always be possible, for example where UAVs are deployed to monitor ground sensors in a forest. The energy transfer efficiency is influenced by both distance and the presence of obstacles (line-of-sight vs no-line-of-sight). Our approach offers flexibility to address both these issues. We can position the airborne energy sources in a way that would minimize this distance and improve line-of-sight RF links thus increase energy transfer efficiency. Moreover, our work is the first to consider the optimal positioning of $tUAV$s which maximizes the total received energy in the $rUAV$s.

Our contributions in this paper are as follows: (i) we propose an UAV re-charging architecture using wireless power transfer from carefully positioned, airborne, stationary energy sources that provide power to the UAVs without disrupting their trajectories, (ii) we provide a mathematical model to derive optimal placement of the energy sources to maximize the total received energy in the system, and (iii) we consider a specific scenario of two $rUAV$s moving along a linear trajectory servicing ground nodes stationed within a square region and use our model to determine the optimal locations for two $tUAV$s. From our solutions to the optimal placement problem, we observed that for this specific scenario, the optimal placement of an even number of energy sources will also result in fairness in terms of equal amount of received energy by all $rUAV$s. However, we found that if we used an odd number of energy sources, either the fairness could be achieved or the total amount of received energy could be maximized, but not both at the same time. Our numerical results revealed that placing the charging nodes at the suggested optimal locations resulted in significant power gain compared to non-optimal placements.

The rest of the paper is organized as follows. Section \unskip~\ref{sec:uca} presents our proposed UAV recharging architecture. We first present a general case of any number of $tUAV$s and $rUAV$s, followed by a specific case of two $tUAV$s and two $rUAV$s. We solve the specific case of energy source placement in Section \unskip~\ref{sec:ssc}. We provide implications of our solutions in Section \unskip~\ref{ram}, ending our paper with some numerical results and conclusion in Sections \unskip~\ref{num} and \unskip~\ref{conclu}.

\bgroup
\fixFloatSize{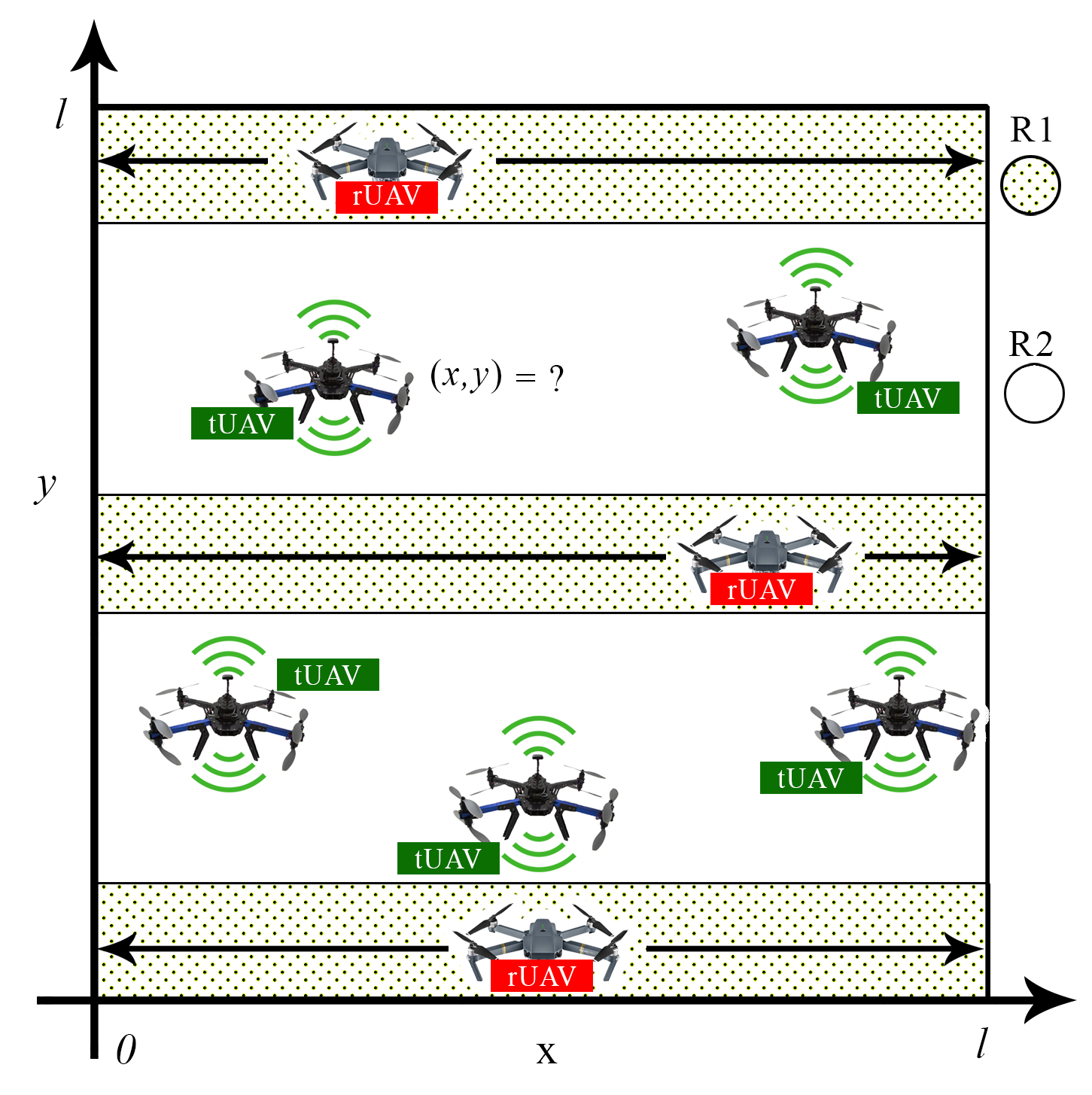}
\begin{figure}[!tbp]
\centering \makeatletter\IfFileExists{Fig1_NEW.jpg}{\includegraphics[width=8cm,height=6cm]{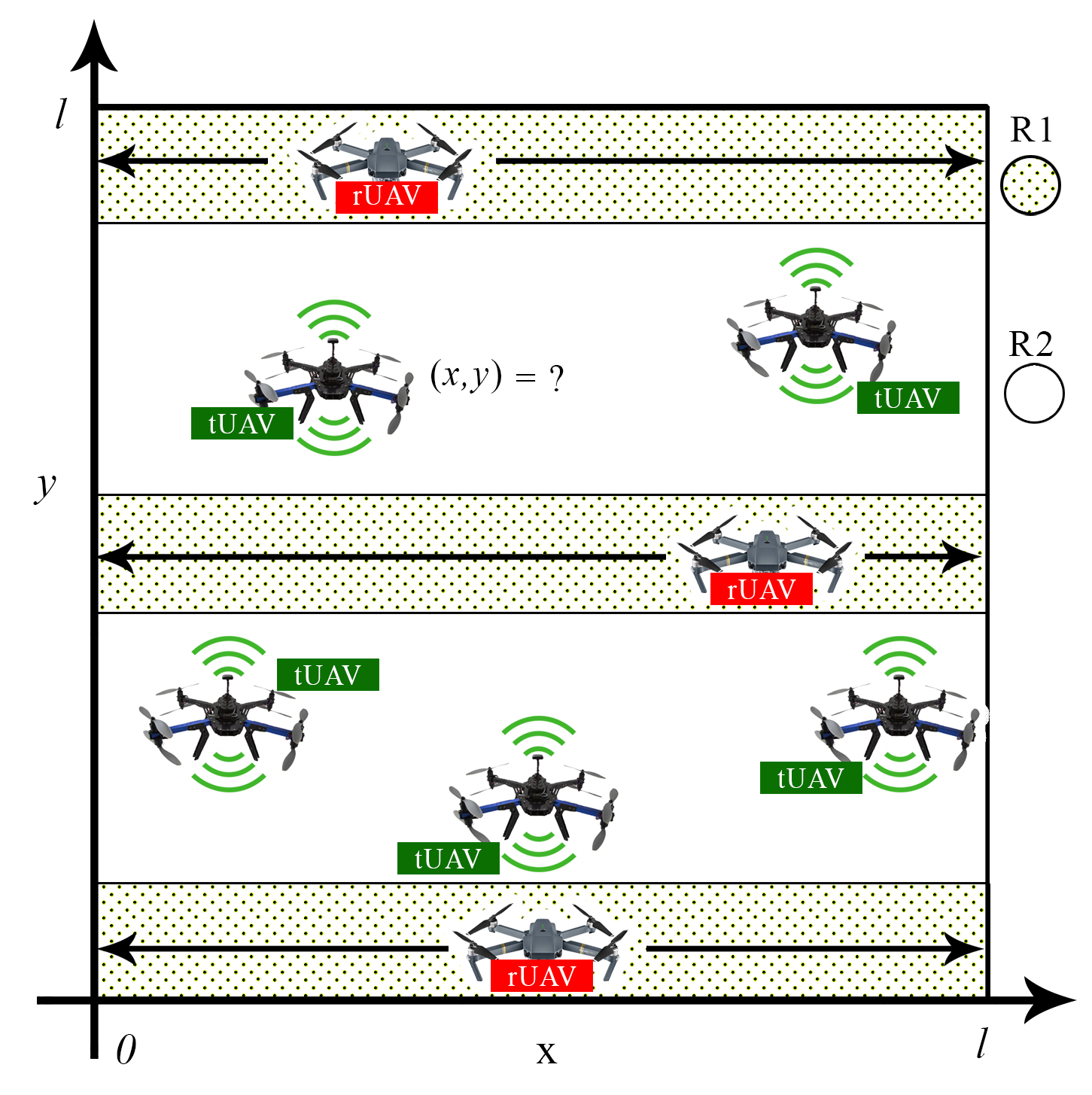}}
\makeatother 
\caption{The system model for flying base station recharging. The $rUAV$s represent the power-receiver $UAV$s, and the $tUAV$s represent the RF energy sources. Only the $(x,y)$ coordinates of the $tUAV$s are shown, since these are placed at the same height as those of the $rUAV$s.}
\label{figure-14f4abc00bef73fd1e0fb3b53d0d4368}
\end{figure}
\egroup

\section{UAV Recharging Architecture: System Model}
\label{sec:uca}
 Our UAV charging architecture is shown in Figure \unskip~\ref{figure-14f4abc00bef73fd1e0fb3b53d0d4368}, which is used for charging a number of cruising UAVs that fly back and forth with a linear trajectory over a square area of side length $l$. The trajectories of the cruising UAVs can be of any form, e.g., of geometric form (circular, linear), or can be of other forms, as shown in \unskip~\cite{7974285, flyingdronebs}. In our work, we have assumed a linear trajectory for the cruising UAVs. These are the RF energy receiver UAVs- the $rUAV$s, and fly back and forth on a path parallel to the horizontal axis of the square area, with a constant speed $V$. The $rUAV$s harvest energy from the received RF signals while in service, from airborne, dedicated energy sources. We assume that these energy sources are specialized UAVs, equipped with wireless power transmitters, and are referred to as transmitter UAVs- $tUAV$s. The $tUAV$s are placed at fixed locations (i.e., non-mobile) over this area with their $(x,y)$ coordinates given by $(x_1,y_1)$ \& $(x_2,y_2)$, etc., and their $z$ coordinate (the height) is the same as those of the $rUAV$s. The heights of the $tUAV$s and the $rUAV$s being same improves the amount of received RF power, which we can adjust since we are using airborne energy sources, as opposed to using terrestrial energy sources with non-adjustable heights. The $tUAV$s are assumed to have a wire-line connection to the ground for a constant power supply \unskip~\cite{wireline_drone}. In order to avoid the chance of collision, the $tUAV$s must be placed outside the collision zone $R1$, anywhere in zone $R2$ as shown in Figure \unskip~\ref{figure-14f4abc00bef73fd1e0fb3b53d0d4368}. By careful positioning of the $tUAV$s in terms of their $(x, y)$ coordinates, this architecture aims to maximize the received energy by the $rUAV$s during their flight time to travel one side of the square, achieving service continuity by the $rUAV$s without disrupting their trajectory. We provide a general model for this architecture next.
 
 Time taken to travel one side length of the square area by a $rUAV$ is given by $T={l}/{V}$. Locations of the $rUAV_1$ and $rUAV_2$ that are on the parallel edges of the square at time $t$ is given by $(Vt,0)$ and $(Vt,l)$ respectively, for $t\in[0,T]$. The received power of far-field RF transmission attenuates as per the  reciprocal of the squared distance between the transmitter and the receiver. Therefore, the harvested RF power ($P_R$) at the receiver can be calculated using Frii's free space propagation model \unskip~\cite{friis-2} as:
\begin{equation}\label{eqfriis}
 P_R = \frac{P_TG_TG_R\lambda^{2}}{(4\pi R)^{2}}
\end{equation}
where $P_T$ is the transmit power, $G_T$ and $G_R$ are the antenna gains of the transmitter and the receiver, $\lambda$ is the power transfer wavelength, and $R$ is the distance between the transmitter and the receiver. Without the loss of generality, we can say that the received power varies inversely with the \textit{square of the distance} between the transmitter and the receiver, which our model is based on. In Section \unskip~\ref{num}, we use specific values of the other parameters of Frii's equation to estimate received power. Distance of $rUAV_1$ from $tUAV_1$ at time $t$ is $ \sqrt{(Vt-x_1)^2+y_1^2}$. So, energy received by $rUAV_1$ from the $tUAV_1$ over $[0,T]$ is 

\begin{equation}
\propto \int_0^T \frac{dt}{(Vt-x_1)^2+y_1^2}.
\end{equation}

For a general $rUAV$ path $(x(t), y(t)), \phantom{-}0 \le t \le T$, i.e., the $rUAV$ located anywhere in the considered area, energy received by an $rUAV$ from a $tUAV$ located at $(x_1,y_1)$ is

\begin{equation}
\propto \int_0^T \frac{dt}{(x(t)-x_1)^2+(y(t)-y_1)^2}. 
\end{equation}
Let $E_{rUAV_k,tUAV_j}$ be the energy received by $rUAV_k$ from $tUAV_j$ over time $0 \le t \le T$. Then the total energy received by $rUAV_k$ is:
\begin{equation}
E_k = \sum_{j}   E_{rUAV_k,tUAV_j} \propto \sum_{j} \int_0^T \frac{dt}{(x_k(t)-x_j)^2+(y_k(t)-y_j)^2}. 
\end{equation}
The total energy received by all $rUAV$s from all $tUAV$s is given by:
 
 \begin{equation}
E_{total} = \sum_{k} E_{k} = \sum_{k}\sum_{j} E_{rUAV_k, tUAV_j}
\end{equation}
where $(x_k(t),y_k(t))$ is the flight path of $rUAV_k$ for $0 \le t \le T$ and $(x_j,y_j)$ is the $j^{th}$ transmitter UAV's ($tUAV_j$) location. The $E_{total}$ can also be calculated by summing up the given energy by all $tUAV$s to all $rUAV$s, as:

\begin{equation}
E_{total} = \sum_{j} \sum_{k} E_{rUAV_k, tUAV_j}
\end{equation}
where $\sum_{k} E_{rUAV_k, tUAV_j}$ is the energy provided by $tUAV_j$ to all $rUAV$s. In order to gain an insight into solving the energy source placement problem, we focus on a specific case of \textit{two transmitters} and \textit{two receivers} next. %Our objective is to study the placement of the two transmitter $UAV$s that will result in the maximum received energy in the two receiver $UAV$s over a given time period.  

\subsection{The Case of Two $tUAV$s and Two $rUAV$s}
In this section, we consider a scenario where two $tUAV$s ($tUAV_1$ and $tUAV_2$) are placed at locations $(a_1,b_1)$ \& $(a_2,b_2)$, at the same height level of two $rUAV$s ($rUAV_1$, and $rUAV_2$). The $rUAV$s fly back and forth over straight-line paths of two parallel edges of the square, separated from each other with a distance $l$ which is the length of the square. Note that the paths of the two $rUAV$s are given by $(x_1(t),y_1(t))=(Vt,0)$, and $(x_2(t),y_2(t))=(Vt,l)$ for $0 \le t \le T$. Total energy received by $rUAV_1$ is given by

\begin{equation}
\begin{aligned}
E_1 \propto \int_0^T \frac{dt}{(x_1(t)-a_1)^2+(y_1(t)-b_1)^2} +\\
\int_0^T \frac{dt}{(x_1(t)-a_2)^2+(y_1(t)-b_2)^2}. 
\end{aligned}
\end{equation}
Replacing the path position values of $rUAV_1$, we get
\begin{equation}
\begin{aligned}
E_1 \propto \int_0^T \frac{dt}{(Vt-a_1)^2+b_1^2} +\\
\int_0^T \frac{dt}{(Vt-a_2)^2+b_2^2}. 
\end{aligned}
\end{equation}
Similarly,
\begin{equation}
\begin{aligned}
E_2 \propto \int_0^T \frac{dt}{(Vt-a_1)^2+(l-b_1)^2} +\\
\int_0^T \frac{dt}{(Vt-a_2)^2+(l-b_2)^2}. 
\end{aligned}
\end{equation}\\
So, our objective is to maximize $E_1+E_2$, or equivalently

\begin{equation*}
P:\max_{a_1,b_1,a_2,b_2} \quad  E_1+E_2
\end{equation*}
\begin{equation*}
\text{s.t.} \quad  0 \le a_j \le l \; \quad j=1,2
\end{equation*}
\begin{equation}
\varepsilon \le b_j \le l -\varepsilon \; \quad j=1,2
\end{equation}
Here, $\varepsilon \in (0,l/2)$ is the width of the region $R1$ as in Figure \unskip~\ref{figure-14f4abc00bef73fd1e0fb3b53d0d4368}, representing the collision area width of each $rUAV$ within which no $tUAV$s are to be placed. %Note that to make $E_1=E2$ we cannot have both the tUAVs in the same side of $y=l/2$ as otherwise one rUAV will receive more power than the other as it will be closer to the transmitter. As such, we add in the constraints $b_1 \ge l/2, b_2 \le l/2$, without loss of generality.
\bgroup
\fixFloatSize{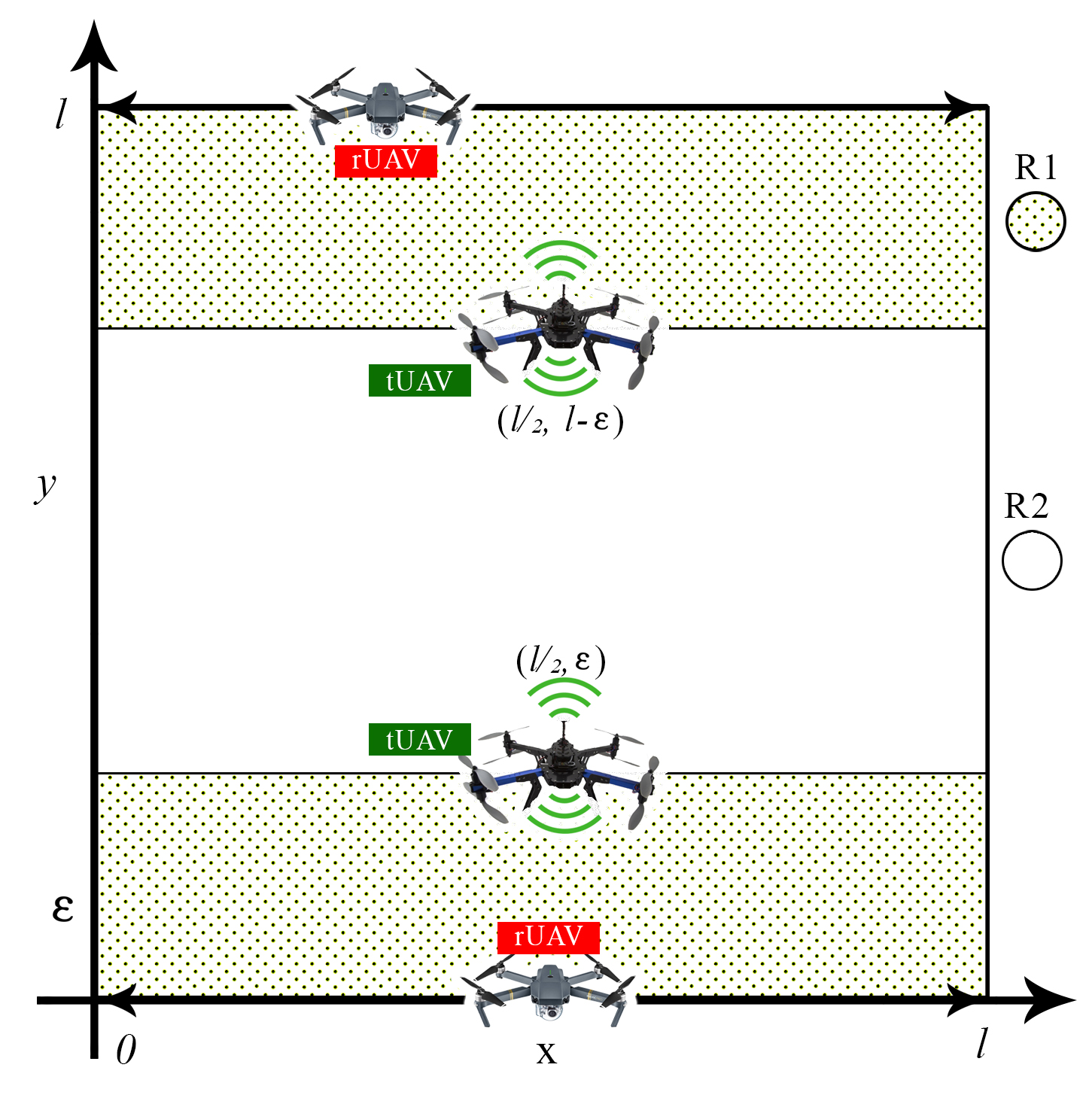}
\begin{figure}[!tbp]
\centering \makeatletter\IfFileExists{Fig2_NEW.jpg}{\includegraphics[width=8cm,height=6cm]{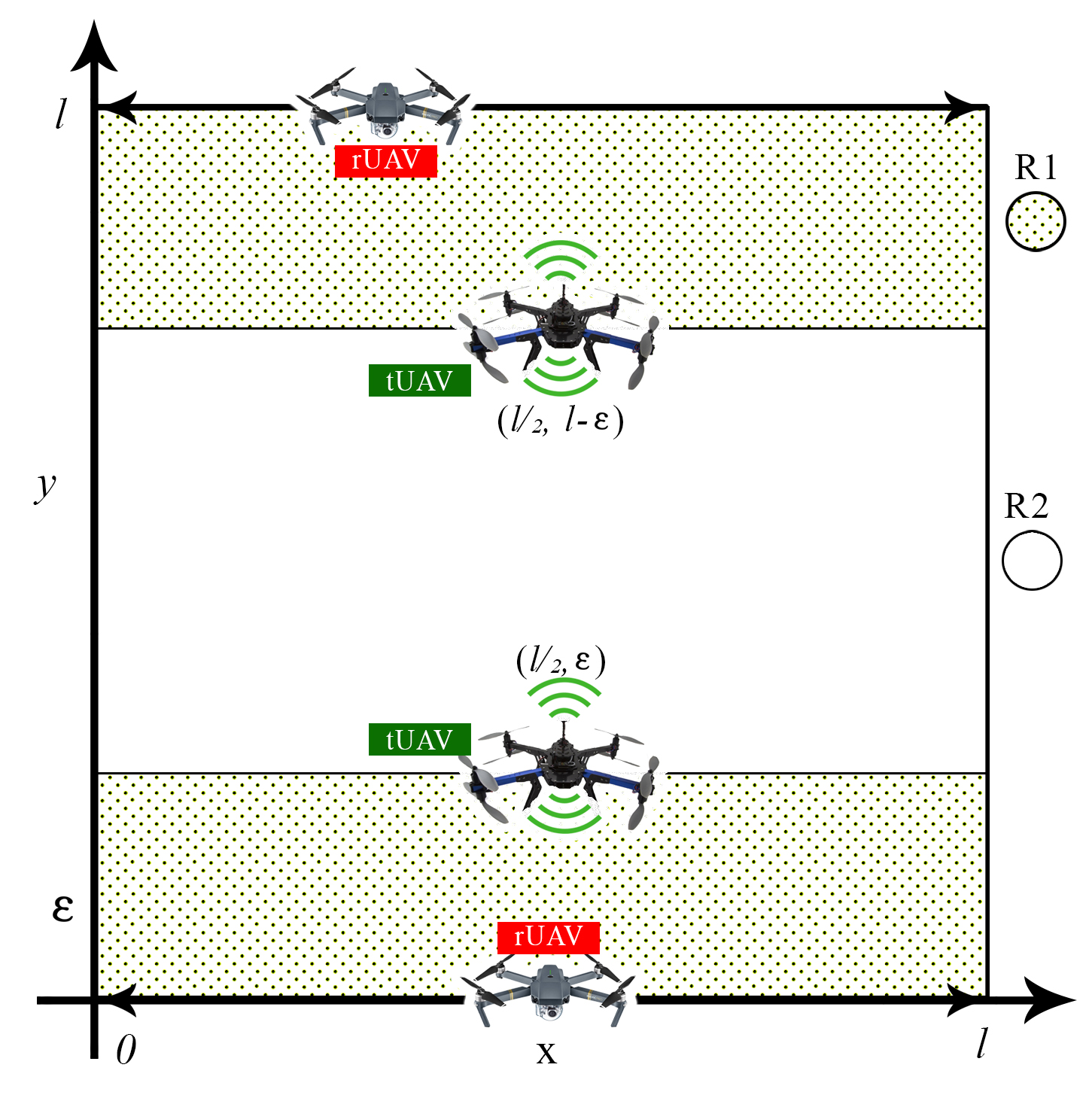}}
\makeatother 
\caption{{Optimal placement of two $tUAV$s to recharge two $rUAV$s.}}
\label{figure-fig2}
\end{figure}
\egroup

\section{Optimal Solution to Problem P}
\label{sec:ssc}
In this section, we solve the $tUAV$ placement problem for the specific case of two $tUAV$s and two $rUAV$s as per the description in the previous section, with the restriction of not placing any $tUAV$ in the region $R1$ of each $rUAV$. We show that the optimal placement of the $tUAV$s are one transmitter on each boundary of the restricted zones of each $rUAV$, and in the middle of the horizontal axis of the zone boundary which is shown in Figure \unskip~\ref{figure-fig2}. Our main result is the following Theorem.

\begin{theorem}\label{thm1}
The physically unique solution to Problem $(P)$ is
\begin{center}
$(a_1, b_1) = (l/2,\varepsilon)$, $(a_2, b_2) = (l/2,l-\varepsilon)$.       
\end{center}
\end{theorem}

In order to help prove the theorem, we state a series of useful Lemmas that we prove in the Appendix.

\begin{lemma}\label{lma1}
Let $c:[0,l] \rightarrow \mathbb{R}$ \textrm{be a (strictly) concave function.}\\
Let $g:[0,l]\rightarrow \mathbb{R}, \quad g(x)=c(x)+c(l-x) \quad \forall x \in [0,l].$\\
Then $g$ \textrm{is (strictly) maximized at} $x=\frac{l}{2}.$
\begin{proof}
See Appendix A. 
\end{proof}
\end{lemma}
\begin{lemma}\label{lma2}
Let $f:[\varepsilon, l-\varepsilon] \rightarrow \mathbb{R}$ \textrm{be (strictly) convex, where } $\varepsilon \lt l/2$. \\
Let $h:[\varepsilon, l-\varepsilon] \rightarrow \mathbb{R}, \quad h(x)=f(x)+f(l-x) \quad \forall x \in [\varepsilon,l-\varepsilon].$\\
Then $h$ \textrm{is (strictly) maximized at the endpoints (i.e., at $x=\varepsilon$ \quad and $x=l-\varepsilon$)}. 
\begin{proof}
See Appendix B. 
\end{proof}
\end{lemma}
%Then note $E_1+E_2$ is of the form of $g$ of Lemma \unskip~\ref{lma1} as a function of $a_1$ (and of $a_2$), and is of the form of $h$ in Lemma \unskip~\ref{lma2} as a function of $b_1$ (and $b_2$).

\iffalse
FIG 3
\bgroup
\fixFloatSize{images/Fig3.jpg}
\begin{figure}[!tbp]
\centering \makeatletter\IfFileExists{images/Fig3.jpg}{\includegraphics{images/Fig3.jpg}}
\makeatother 
\caption{{Optimum placement of one Charger UAV dictates the placement of the other Charger UAV, to be symmetrical opposite of the middle line}}
\label{figure-fig3}
\end{figure}
\egroup
\fi

Below we provide the proof of Theorem \ref{thm1}.
\begin{proof}
We have to maximize total energy received by the $rUAV$s from $tUAV_1$ and $tUAV_2$, which is given by
\begin{equation*}
\begin{aligned}
 \textrm{Total Energy = (Energy provided by } tUAV_1 ) + \\
 (\textrm{Energy provided by } tUAV_2).
\end{aligned}
\end{equation*}

Energy provided by $tUAV_j$ is  
\begin{equation*}
\begin{aligned}
= \int_0^T \frac{dt}{(Vt-a_j)^2+b_j^2} & +
 \int_0^T \frac{dt}{(Vt-a_j)^2+(l-b_j)^2} \\
=  \phi (a_j,b_j) + & \phi (a_j, l-b_j), \\
\end{aligned}
\end{equation*}

where $\phi(a,b) := \int_0^T \frac{dt}{(Vt-a)^2+b^2}$, $\quad  0 \le a \le l,\quad \varepsilon \le b \le l -\varepsilon$
and so,
\begin{equation*}
E_{total} \propto \phi(a_1,b_1) + \phi(a_1, l-b_1) + \phi(a_2,b_2) + \phi(a_2, l-b_2).
\end{equation*}
Since total energy received are additions of energy contributed by each $tUAV$, which come from the same function $\phi(a,b) + \phi(a, l-b)$ with independent values of $(a,b)$, it suffices to find how to maximize $\phi(a,b)+\phi(a,l-b)$ for $ 0 \le a \le l, \text{and } \varepsilon \le b \le l -\varepsilon$.
Let $F(a,b) = \phi(a,b) + \phi(a, l-b)$. Note that $F(a,b)$ is proportional to total energy received by both $rUAV$s in travelling one side length of the square, from one $tUAV$ located at $(a,b)$. We want to maximize $F$ over $(a,b)$.

We prove the following two properties of $F(a,b)$:

\begin{Properties}
  \item  $\argmax\limits_{a \in [0,l]} F(a,b)= \frac{l}{2} \quad \forall b \in [\varepsilon, l - \varepsilon]$
  
 We have
\begin{align*}
\phi(a,b)&=\int_0^T \frac{dt}{(Vt-a)^2+b^2}\\
&= \frac{1}{Vb}\Big (\tan^{-1}\Big(\frac{a}{b}\Big) + \tan^{-1}\Big(\frac{l-a}{b} \Big) \Big ).
\end{align*}      
Recall, $F(a,b) = \phi(a,b) + \phi(a, l-b)$. Hence to show $\argmax\limits_{a \in [0,l]} F(a,b)= \frac{l}{2} \quad \forall b \in [\varepsilon, l - \varepsilon]$, it suffices to show that 
 
  \begin{equation}\label{eqpr1}
 \argmax\limits_{a \in [0,l]}\phi(a,b)= \frac{l}{2} \quad \forall b \in [\varepsilon, l - \varepsilon].    
  \end{equation}
 
  Because $ \forall b \in [\varepsilon, l - \varepsilon]$ we have $\frac{1}{Vb}\gt 0$ and $\tan^{-1}\Big (\frac{a}{b} \Big)$ is strictly concave in $a$ for $a\in [0,l]$, Lemma \unskip~\ref{lma1} implies the desired result of Equation \unskip~\ref{eqpr1}.\\
  
\item  $\argmax\limits_{b \in [\varepsilon, l-\varepsilon]} F(a,b) = \{\varepsilon, l-\varepsilon\} \quad \forall a \in [0, l]$
 
We first show for any constant $k\gt0$, $\frac{1}{Vb}\tan^{-1}\Big(\frac{k}{b}\Big)$ is strictly convex in $b$ for $b \in[\varepsilon, l-\varepsilon]$. Observe
  
 \begin{equation*}
     \frac{1}{Vb}\tan^{-1}\Big(\frac{k}{b}\Big) = f(g(b))
 \end{equation*}
 where $f(x)=\frac{x}{V}\tan^{-1}\Big(kx\Big)$ and $g(b)=\frac{1}{Vb}$ (for $b\in[\varepsilon,l-\varepsilon]$). Since if $f$ is a convex and strictly increasing function, and $g$ is a strictly convex function, then $f(g(b))$ is strictly convex, it suffices to show:
 \begin{itemize}
     \item $g$ is strictly convex. 
     It is clear that $g(b)$ is strictly convex.
     \item $f$ is convex, and strictly increasing. 
     We see that $f$ is strictly increasing because $f(x)$ is a product of two strictly increasing positive functions (for $x \gt 0$).
     To show $f$ is convex, we observe that its second derivative $f^{\prime\prime}(x)$ is $\frac{2k}{V\Big(1+k^2x^2 \Big)^2}\gt 0$ for $k\gt0$. 

Since 
 \begin{align*}
\phi(a,b)& = \frac{1}{Vb}\tan^{-1}\Big(\frac{a}{b}\Big) + \frac{1}{Vb}\tan^{-1}\Big(\frac{l-a}{b} \Big)
\end{align*}  
the above implies that $\phi(a,b)$ is the sum of two strictly convex functions in $b$ for fixed $a \in (0,l)$. For $a=0 \text{ or } l$, $\phi(a,b) = \frac{1}{Vb}\tan^{-1}\Big(\frac{l}{b}\Big)$, which is also a strictly convex function of $b$. 
Therefore, for any $a \in [0,l]$, $\phi(a,b)$ is strictly convex in $b$. Since $F(a,b) = \phi(a,b) + \phi(a, l-b)$, Lemma \unskip~\ref{lma2} now implies Property 2. 
 \end{itemize}
 \end{Properties}
These properties prove $F(a,b)$ is maximized precisely at $(l/2,\varepsilon)$ and $(l/2,l-\varepsilon)$. Recall $F(a,b)$ is proportional to total energy received by both $rUAV$s in travelling one side length of the square, from one $tUAV$ located at $(a,b)$. Total received energy is maximized if we place the (one) $tUAV$ at either of these two points. As such, if we have two $tUAV$s, we need to place one $tUAV$ at $(l/2,\varepsilon)$ and the other $tUAV$ at $(l/2,l-\varepsilon)$ for the total received energy to be maximized, as total energy is the sum of energy received by the $rUAV$s from both of the $tUAV$s.

Thus, Theorem \unskip~\ref{thm1} is proved.
\end{proof}

\section{Ramifications of the Model}
\label{ram}
Figure \unskip~\ref{figure-fab} shows the values of $F(a,b)$, which is proportional to the received energy by two $rUAV$s for various placements of one $tUAV$ within zone $R2$. We observe that placing the $tUAV$ anywhere but at either of the two mid-points of the boundaries of $R1$ and $R2$ results in lower values of $F(a,b)$. As such, placing the $tUAV$s at these optimal locations will result in maximum total received energy by the two $rUAV$s. If we place an even number of $tUAV$s by equally distributing these at these two locations, we will achieve an equal amount of received energy in each $rUAV$ (fairness). However, if we have an odd number of $tUAV$s, there will be an imbalance of energy received individually by the $rUAV$s, since after distributing the $tUAV$s equally in these two locations, if we place the leftover $tUAV$ in one of these two locations, received total energy will be maximized, however, this will also mean that the $rUAV$ closer to the last $tUAV$ will receive more energy than the other (unfair). If we place the last $tUAV$ in the middle position from both the $rUAV$s, they will receive an equal amount of energy but the total received energy will not be maximized since the last $tUAV$ is at a non-optimal location. Thus, we make the following observations:
\iffalse
\bgroup
\fixFloatSize{images/Fab_V10ms_l80_e5.jpg}
\begin{figure}[!tbp]
\centering \makeatletter\IfFileExists{images/Fab_V10ms_l80_e5.jpg}{\includegraphics[width=8cm,height=6cm]{images/Fab_V10ms_l80_e5.jpg}}
\makeatother 
\caption{$F(a,b)$ for $l=80m, \varepsilon=5m$, $V=10m/s$.}
\label{figure-fab}
\end{figure}
\egroup
\fi

\begin{figure}
 \centering % centering figure
 %\scalebox{0.9} % rescale the figure by a factor of 0.8
 {\includegraphics[width=8cm,height=5cm]{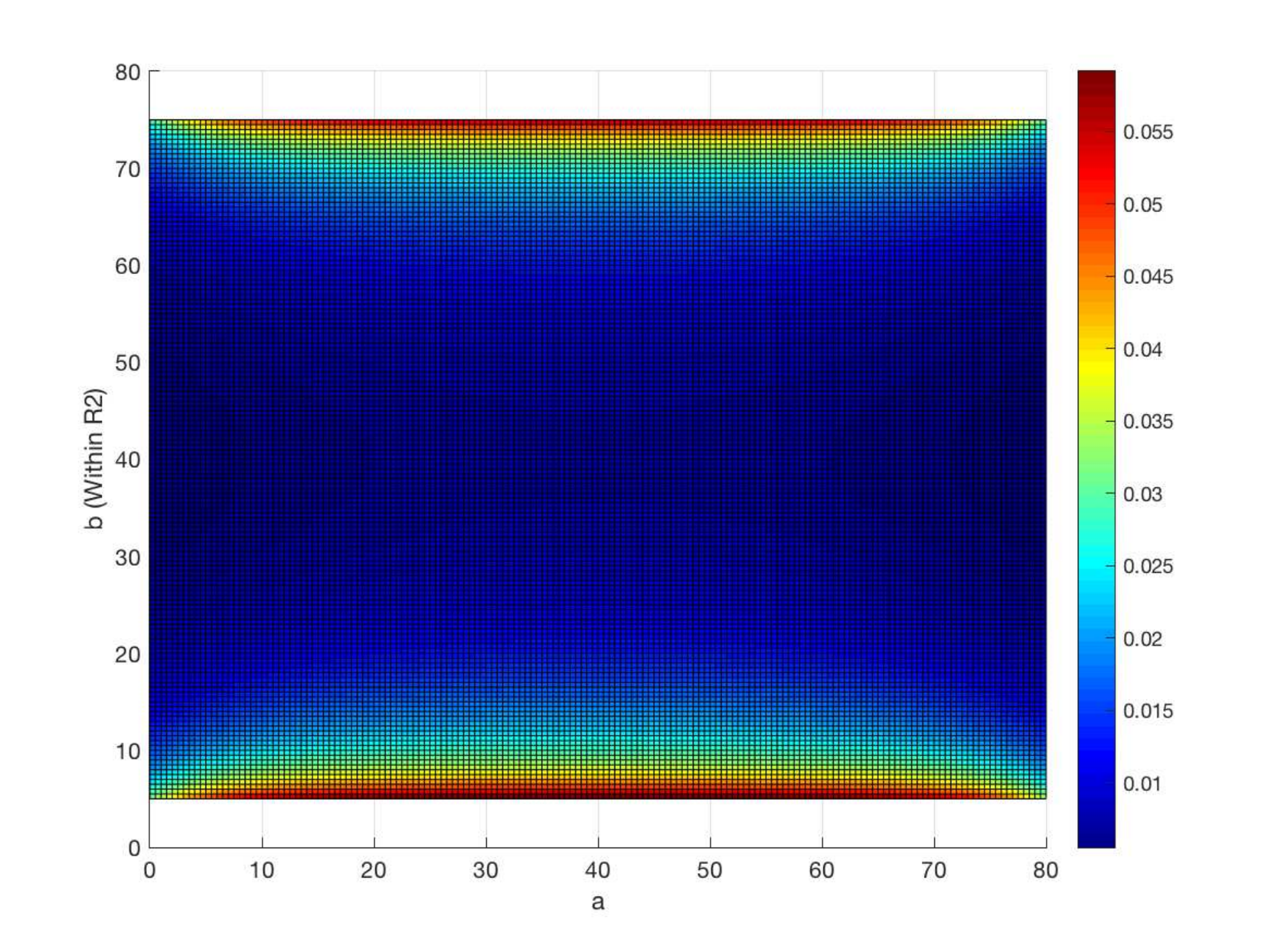}} % importing figure
 \caption{$F(a,b)$ within $R2$, for $l=80m, \varepsilon=5m$, $V=10m/s$. The two $rUAV$s are flying over the opposite horizontal axes of the square area.}
 \label{figure-fab} % labeling to refer it inside the text
\end{figure}

\begin{Observation}
To recharge two $rUAVs$ with an even number of $tUAVs$, it is possible to place the $tUAVs$ in such a way that maximizes $E_{total}$, i.e., $E_1+E_2$, and also achieves fairness, i.e., $E_1=E_2$.
\end{Observation}

\begin{Observation}
To recharge two $rUAVs$ with an odd number of $tUAVs$, it is  possible to achieve either maximized $E_{total}$, or fairness ($E_1=E_2$), but not both at the same time. \end{Observation}

\begin{Observation}
The optimal placement locations are valid for recharging two $rUAVs$ by \textbf{any number of $tUAVs$}, i.e., not only by two $tUAVs$, since new $tUAVs$ contribute to the total energy in an additive manner. 
\end{Observation}

\section{Numerical Results}
\label{num}
In order to gain an insight into the average power that the optimal placement of \textbf{one $tUAV$} can provide to the \textbf{two $rUAV$s} compared to the non-optimal placements, we report the numerical results for a similar scenario as used in our model but with one $tUAV$. The parameter values are listed in Table \unskip~\ref{tab:para}. The calculation is based on Equation ~\unskip\ref{eqfriis}, however, we replaced $\lambda$ with $\frac{c}{f}$ where $c$ is the speed of light, and $f$ is the frequency.  

For the $tUAV$ power transmission frequency, we have used $433$ MHz, which belongs to the non-licensed ISM band. This frequency is commonly found to be generated by garage door openers, however, we use this in our numerical experiments for the RF power transmission. Using this frequency in a remote location should not cause interference with other devices. Note that the commercially available RF transmitters that are used for charging low-power devices use higher frequencies, for example, \textit{Powercaster} transmitters \unskip~\cite{powercaster} use $915$ mHz. Our requirement of charging UAVs demanding higher power, the $tUAV$s need to transmit power at lower frequencies, since lower frequencies result in higher received power. Moreover, due to the significant reduction of RF power at the receiver compared to the transmitted power, we also need the $tUAV$ to transmit at a higher power, which we have taken to be $1$ kW. This can be justified by our assumption that the $tUAV$s have ground power supply connections, thus it can transmit at this level. Results discussed below assumes full energy conversion efficiency. 
\begin{figure}
\centering % centering figure
 %\scalebox{0.9} % rescale the figure by a factor of 0.8
 {\includegraphics[width=8cm,height=5cm]{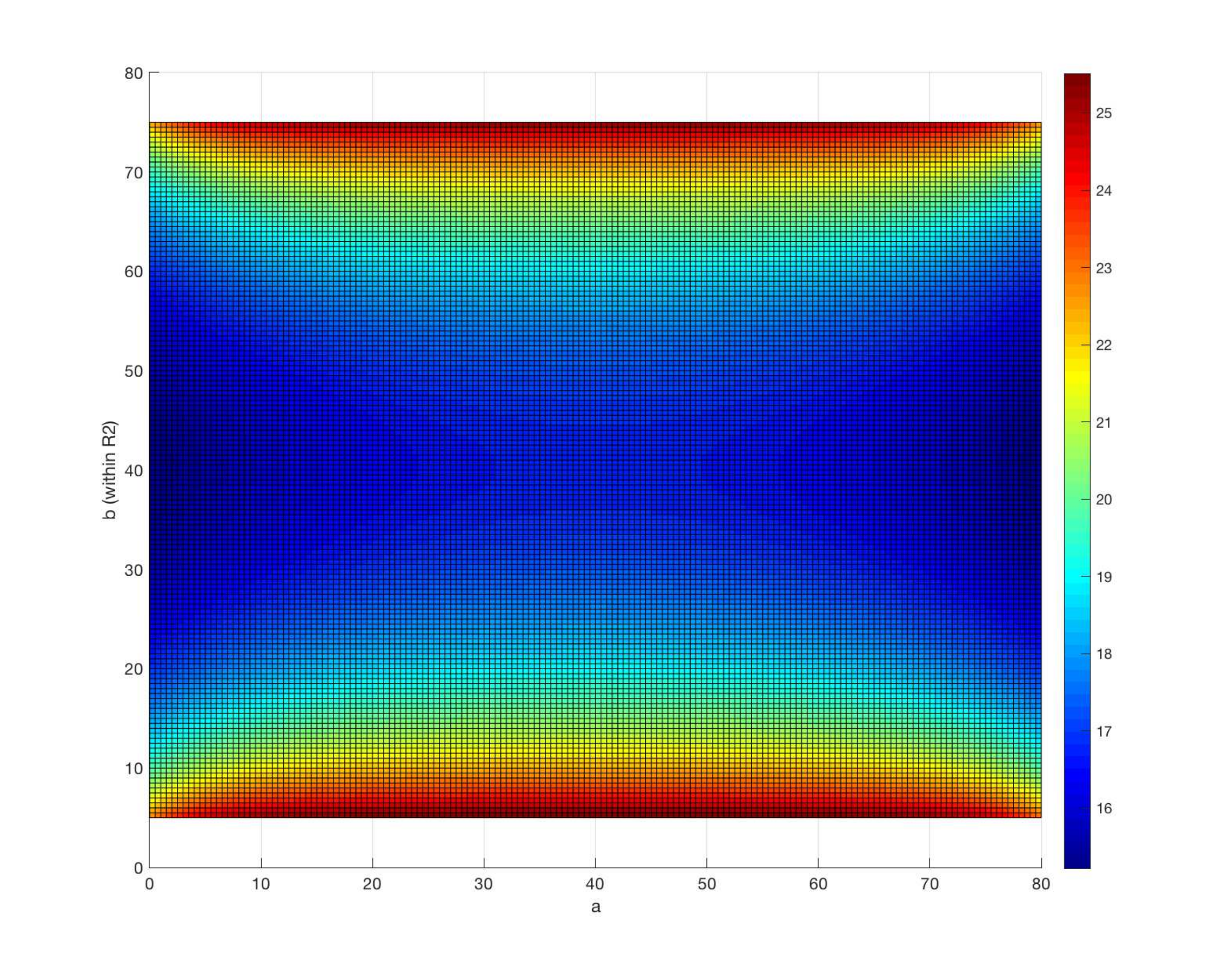}} % importing figure
 \caption{Average power (dBm) received by two $rUAV$s from one $tUAV$ located at $(a,b)$, over the time interval $[0,T]$. The two $rUAV$s are flying over the opposite horizontal axes of the square area.}
 %, f=433 mHz, P_t=1000 W$).}
 \label{fig:avg_power_DBM_433_1K} % labeling to refer it inside the text
\end{figure}
%26 jan fig below

\iffalse quoted out the 915 mhz 57 dbm figure from 26 January ******
\begin{figure}
\centering % centering figure
 %\scalebox{0.9} % rescale the figure by a factor of 0.8
 {\includegraphics[width=8cm,height=5cm]{images/avg_power_dbm_26Jan.eps}} % importing figure
 \caption{Average Power (dBm) received by Two $rUAV$s from one $tUAV$ located at $(a,b), f=915 mHz, P_t=501 W (57dBm)$.}
 \label{fig:avg_power_DBM} % labeling to refer it inside the text
\end{figure}
\fi

\begin{table}
\begin{center}
\caption{Parameter Values}
\label{tab:para} 
\begin{tabular}{c|c}
\hline
 Parameter & Value  \\
 \hline
 $f$ & $433$ MHz  \\ 
 $c$ & $299792458$ m/s\\
 $P_t$ &  $1$ kW \\
 $G_r,G_t$ & $6$ dBi\\
 $V$ & $10$ m/s\\
 $l$ & $80$ m\\
 $\varepsilon$ & $5$ m\\
 \hline
\end{tabular}
\end{center}
\end{table}

\begin{figure}
\centering % centering figure
 %\scalebox{0.9} % rescale the figure by a factor of 0.8
 {\includegraphics[width=8cm,height=5cm]{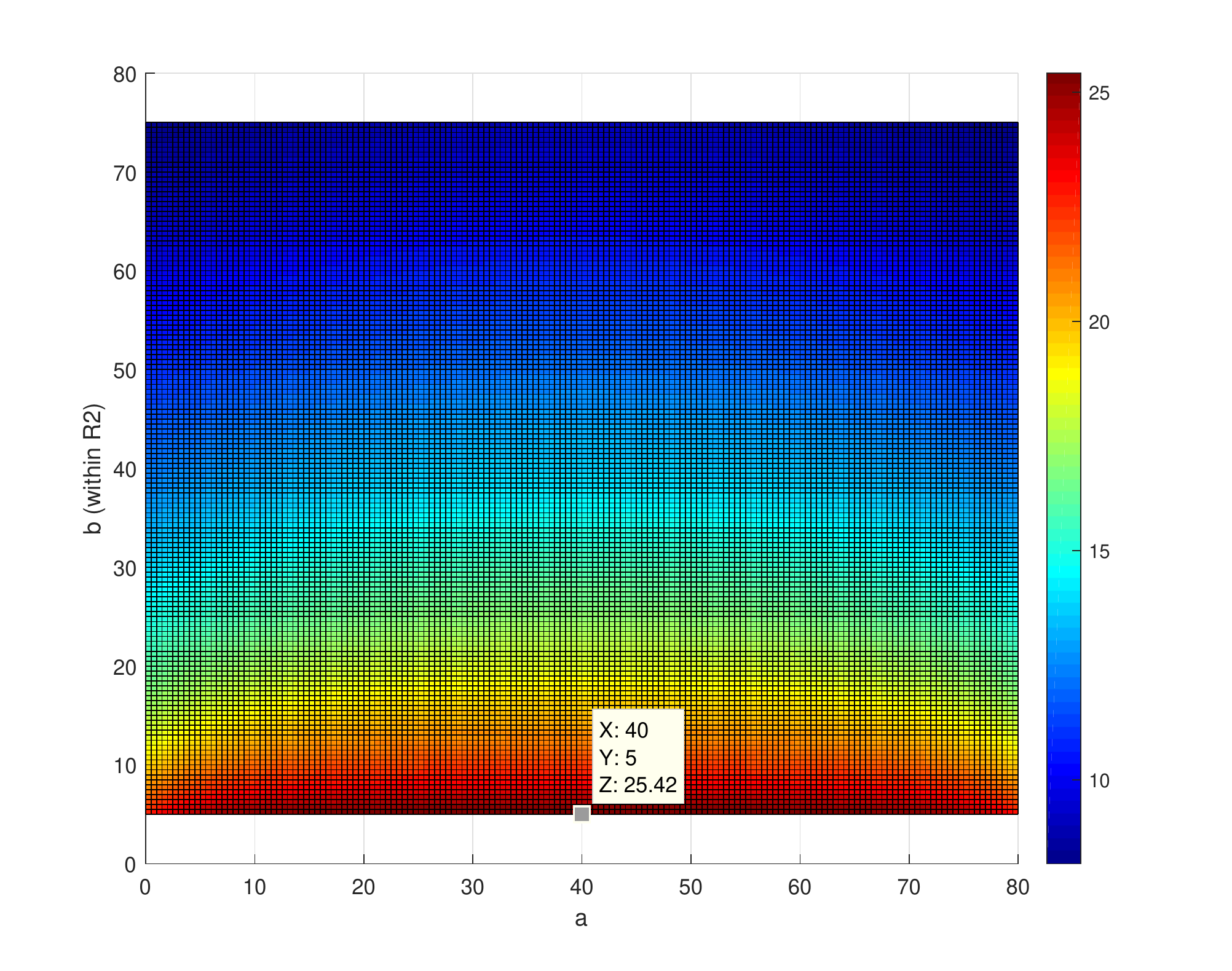}} % importing figure
 \caption{Average power (dBm) received by one $rUAV$s from one $tUAV$ located at $(a,b)$, over the time interval $[0,T]$. The $rUAV$ is flying over the lower horizontal axis of the square area.}
 %, f=433 mHz, P_t=1000 W$).}
 \label{fig:avg_power_1uav} % labeling to refer it inside the text
\end{figure}

Figure \unskip~\ref{fig:avg_power_DBM_433_1K} shows the average power over the time taken to traverse one side of the square (for the $rUAV$) in dBm, received by the two $rUAV$s from one $tUAV$ placed at different $(x,y)$ coordinates within the safe placement zone $R2$ of the square area. The optimal placement of the $tUAV$ as per our model, which is the middle of either of the boundary of $R1$ and $R2$ (the mid point of the horizontal lines of the coloured zone) resulted in the maximum average received power of $25.5121$ dBm by the two $rUAV$s. If the $tUAV$ is placed at the middle of the area, i.e, at $(40,40)$, the average received power became $16.7425$ dBm. Placing the $tUAV$ at the middle of the vertical axes, i.e., at $(0, 40)$, or $(80,40)$ resulted in the average received power of $15.2233$ dBm. As we can see, the optimal placement of the $tUAV$ resulted in a power gain of $8.7696$ dBm and $10.2888$ dBm over the reported two \textit{non-optimal} placements. 

In order to observe the level of power received by one $rUAV$ from one power source, in Figure \unskip~\ref{fig:avg_power_1uav} we have reported the average power over the time taken to traverse one side of the square when the $tUAV$ is placed at different $(x,y)$ coordinates. In this case, the average power received by the $rUAV$ is found to be $25.4152$ dBm when the $tUAV$ is placed at the optimal location which is at $(40,5)$ in this case, and $12.2130$ dBm when it is located at the middle of the vertical axes. A gain of $13.2022$ dBm was achieved in this case by placing the $tUAV$ at the optimal location. 

Clearly the optimal placement of the $tUAV$ must be used in our considered scenario for the best outcome of the energy transmission system, however, received level of energy harvested from RF signals is still quite low, and will require multiple dedicated RF sources to power the UAVs \unskip~\cite{rfharvesting}. For the fixed-wing drones consuming much less power than the rotary-wing drones since the fixed-wing drones only need energy to move forward but not to keep afloat in air, our system can certainly be considered to extend the operation duration of such drones using multiple $tUAV$s, all placed at (or closer to) the suggested optimal locations. For the placements of multiple $tUAV$s, Section \unskip~\ref{ram} provided some insights.

\section{Conclusion}
\label{conclu}
 This paper considered the problem of in-situ recharging aerial base stations without disrupting their regular trajectory. We proposed a solution that leverages wireless power transfer via carefully positioned airborne but stationary energy sources. We presented a mathematical model for solving the optimal placement of these energy sources so as to maximize the total received power at the UAVs while simultaneously achieving fairness. Our numerical results showed that placing the charging nodes at the suggested optimal locations resulted in significant power gain compared to non-optimal placements. In our future work, we will consider an optimization model that takes the number of $rUAV$s and $tUAV$s as well as their multidimensional trajectories as tuning parameters to find the optimum solution. 
 
 \section*{Appendix} \label{sec:app}
\appendices{A. Proof of Lemma \unskip~\ref{lma1}}\\

Since $c$ is a strictly concave function, therefore $c(l-x)$ is also strictly concave. As such, $g$ being the sum of two strictly concave functions, $g$ is also strictly concave. For $g$ to be (strictly) maximized at $x=l/2$, we have to prove that $g(x) < g(l/2) \quad \forall x \in [0,l]$, and $x \ne l/2$. Let $x \in [0,l]$, and $x \ne l/2$.
%\begin{proof}
We have
\begin{equation*}
\begin{aligned}
g(x) & = c(x) + c(l-x)\\
& = 2\Big(\frac{1}{2} c(x) + \frac{1}{2} c(l-x)\Big)\\
& < 2 c\Big(\frac{1}{2} x + \frac{1}{2} (l-x)\Big) \text{ by strict concavity,}\\
& \text{and since } x \ne l-x\\
& = 2 c\Big(\frac{l}{2}\Big) \\
& = g\Big(\frac{l}{2}\Big).
\end{aligned}
\end{equation*}
Thus, Lemma \unskip~\ref{lma1} is proved.
%\end{proof}

\appendices{B. Proof of Lemma \unskip~\ref{lma2}}\\

Since $f$ is a strictly convex function, therefore $f(l-x)$ is also strictly convex. As such, $h$ being the sum of two strictly convex functions, $h$ is also strictly convex. We have to show that if $\varepsilon \lt x \lt l-\varepsilon$ then $h(x) \lt h(\varepsilon)$ (note $ h(\varepsilon) =  h(l-\varepsilon)$). 
%\begin{proof}
Suppose $\varepsilon\lt x\lt l-\varepsilon$. Then there exists $\theta \in (0,1)$ such that 
\begin{equation*}
x=\theta \varepsilon + (1-\theta)(l-\varepsilon).
\end{equation*}
So,
\begin{equation*}
\begin{aligned}
h(x) & = h\Big( \theta \varepsilon + (1-\theta)(l-\varepsilon) \Big)\\
& < \theta h(\varepsilon) + (l-\theta)h(l-\varepsilon) \text{ by strict convexity,}\\
& \text{ and since } \varepsilon \ne l-\varepsilon\\
& = h(\varepsilon).
\end{aligned}
\end{equation*}

Thus, Lemma \unskip~\ref{lma2} is proved.

%\end{proof}
%\IEEEtriggeratref{13}
\bibliographystyle{IEEEtran}

\bibliography{article}

% Generated by IEEEtran.bst, version: 1.14 (2015/08/26)
\begin{thebibliography}{10}
\providecommand{\url}[1]{#1}
\csname url@samestyle\endcsname
\providecommand{\newblock}{\relax}
\providecommand{\bibinfo}[2]{#2}
\providecommand{\BIBentrySTDinterwordspacing}{\spaceskip=0pt\relax}
\providecommand{\BIBentryALTinterwordstretchfactor}{4}
\providecommand{\BIBentryALTinterwordspacing}{\spaceskip=\fontdimen2\font plus
\BIBentryALTinterwordstretchfactor\fontdimen3\font minus
  \fontdimen4\font\relax}
\providecommand{\BIBforeignlanguage}[2]{{%
\expandafter\ifx\csname l@#1\endcsname\relax
\typeout{** WARNING: IEEEtran.bst: No hyphenation pattern has been}%
\typeout{** loaded for the language `#1'. Using the pattern for}%
\typeout{** the default language instead.}%
\else
\language=\csname l@#1\endcsname
\fi
#2}}
\providecommand{\BIBdecl}{\relax}
\BIBdecl

\bibitem{flyingdronebs}
A.~Fotouhi, M.~Ding, and M.~Hassan, ``{Flying Drone Base Stations for Macro
  Hotspots},'' \emph{IEEE Access}, vol.~6, pp. 19\,530--19\,539, 2018.

\bibitem{dronebsplacement}
L.~Wang, B.~Hu, and S.~Chen, ``{Energy Efficient Placement of a Drone Base
  Station for Minimum Required Transmit Power},'' \emph{IEEE Wireless
  Communications Letters}, pp. 1--1, 2018.

\bibitem{DBLP:journals/corr/abs-1809-01752}
\BIBentryALTinterwordspacing
A.~Fotouhi, H.~Qiang, M.~Ding, M.~Hassan, L.~G. Giordano,
  A.~Garc{\'{\i}}a{-}Rodr{\'{\i}}guez, and J.~Yuan, ``{Survey on {UAV} Cellular
  Communications: Practical Aspects, Standardization Advancements, Regulation,
  and Security Challenges},'' \emph{CoRR}, vol. abs/1809.01752, 2018. [Online].
  Available: \url{http://arxiv.org/abs/1809.01752}
\BIBentrySTDinterwordspacing

\bibitem{7974285}
A.~Fotouhi, M.~Ding, and M.~Hassan, ``{Dynamic Base Station Repositioning to
  Improve Spectral Efficiency of Drone Small Cells},'' in \emph{2017 IEEE 18th
  International Symposium on A World of Wireless, Mobile and Multimedia
  Networks (WoWMoM)}, June 2017, pp. 1--9.

\bibitem{uav_wcomm}
Y.~Zeng, R.~Zhang, and T.~J. Lim, ``{Wireless Communications with Unmanned
  Aerial Vehicles: Opportunities and Challenges},'' \emph{IEEE Communications
  Magazine}, vol.~54, no.~5, pp. 36--42, May 2016.

\bibitem{7994915}
Y.~Li and L.~Cai, ``{UAV-Assisted Dynamic Coverage in a Heterogeneous Cellular
  System},'' \emph{IEEE Network}, vol.~31, no.~4, pp. 56--61, July 2017.

\bibitem{dronebs2}
A.~Fotouhi, ``{Towards Intelligent Flying Base Stations in Future Wireless
  Network},'' in \emph{2017 IEEE 18th International Symposium on A World of
  Wireless, Mobile and Multimedia Networks (WoWMoM)}, June 2017, pp. 1--3.

\bibitem{5gtutorial}
R.~Zhang, ``{UAV Meets Wireless Communication in 5G and Beyond: Main Research
  Challenges and Key Enabling Techniques},'' Tutorial, IEEE Wireless
  Communications and Networking Conference (WCNC), 2018.

\bibitem{UAV5g}
V.~Sharma, K.~Srinivasan, H.~Chao, K.~Hua, and W.~Cheng, ``{Intelligent
  Deployment of UAVs in 5G Heterogeneous Communication Environment for Improved
  Coverage},'' \emph{J. Network and Computer Applications}, vol.~85, pp.
  94--105, 2017.

\bibitem{mabs}
\BIBentryALTinterwordspacing
S.~Enayati, H.~Saeedi, H.~Pishro-Nik, and H.~Yanikomeroglu, ``{Moving Aerial
  Base Station Networks: Stochastic Geometry Analysis and Design
  Perspective},'' 2018 (accessed January 22, 2019). [Online]. Available:
  \url{http://www.ecs.umass.edu/ece/pishro/Papers/ABS.pdf}
\BIBentrySTDinterwordspacing

\bibitem{eurecom}
``{Eurecom PERFUME Project},'' \url{http://www.ercperfume.org/about/},
  (Accessed January 21, 2019).

\bibitem{RF_UAV}
M.~Hua, C.~Li, Y.~Huang, and L.~Yang, ``{Throughput Maximization for
  UAV-enabled Wireless Power Transfer in Relaying System},'' in \emph{2017 9th
  International Conference on Wireless Communications and Signal Processing
  (WCSP)}, Oct 2017, pp. 1--5.

\bibitem{wireline_drone}
\BIBentryALTinterwordspacing
J.~Stewart, ``A startup is setting drones free by tying them to the ground,''
  2018 (accessed March 04, 2019). [Online]. Available:
  \url{https://www.wired.com/story/startup-setting-drones-free-tying-them-to-the-ground/}
\BIBentrySTDinterwordspacing

\bibitem{friis-2}
``{The Friis Equation},'' \url{http://www.antenna-theory.com/basics/friis.php},
  2009-2015 (accessed January 15, 2019).

\bibitem{powercaster}
``{Powercaster Transmitters},''
  \url{https://www.powercastco.com/products/powercaster-transmitter/}, 2018
  (accessed January 14, 2019).

\bibitem{rfharvesting}
X.~Lu, P.~Wang, D.~Niyato, D.~I. Kim, and Z.~Han, ``{Wireless Networks With RF
  Energy Harvesting: A Contemporary Survey},'' \emph{IEEE Communications
  Surveys Tutorials}, vol.~17, no.~2, pp. 757--789, Secondquarter 2015.

\end{thebibliography}

\end{document}